# A Multiple Access Protocol for Multimedia Transmission over Wireless Networks

## Hong Yu and Mohammed Arozullah

Department of Electrical Engineering and Computer Science
Capitol College, Maryland, USA
yhong@capitol-college.edu
Department of Electrical Engineering and Computer Science
The Catholic University of America, Washington DC, USA
arozullah@cua.edu

## ABSTRACT

*This paper develops and evaluates the performance of an advanced multiple access protocol for transmission of full complement of multimedia signals consisting of various combinations of voice, video, data, text and images over wireless networks. The protocol is called Advanced Multiple Access Protocol for Multimedia Transmission (AMAPMT) and is to be used in the Data Link Layer of the protocol stack. The principle of operation of the protocol is presented in a number of logical flow charts. The protocol grants permission to transmit to a source on the basis of a priority scheme that takes into account a time-to-live (TTL) parameter of all the transactions, selectable priorities assigned to all the sources and relevant channel state information (CSI) in this order. Performance of the protocol is evaluated in terms of quality of service parameters like packet loss ratio (PLR), mean packet transfer delay (MPTD) and throughput. Using a simulation model based on an OPNET simulation software package does the evaluation. under various traffic loads with constant distributions with various mean arrival rates and transaction sizes results obtained show that the performance is improved when this priority scheme is used than when it is not used. The results for AMAPMT are compared with that of the best currently available multiple access protocol called Adaptive Request Channel Multiple Access (ARCMA). AMAPMT protocol out performs ARCMA protocol.*

## KEYWORDS

*Network Protocols, Wireless Networks, Multiple Access, Multimedia Transmissions*

## 1. INTRODUCTION

Wireless Networks can be used to transmit multimedia services consisting of Voice, Data, Video, Ftp, and Text. These networks are required to provide desired qualities of service (QoS) to the various media with diverse flow characteristics. For example, packet loss ratio requirement for all loss sensitive services such as email and packet delay requirements for all delay sensitive services as voice are to be satisfied simultaneously and adequately. For a given input traffic load a certain amount of resources (e.g., buffer space and link capacity) are needed to satisfy these QoS requirements. Some of these resources may be scarce and need to be managed well so that maximum amount of input load, with a required QoS requirement, can be accommodated for a given amount of resources. Thus it is required to develop simple and efficient resource management protocols for these networks that can provide better use of network resources.

Three currently available multiple access protocols are considered. These are Packet Reservation Multiple Access (PRMA) [3][17], Distributed Queuing Request Update Multiple Access (DQRUMA) [2] [17] and Adaptive Request Channel Multiple Access (ARCMA) [2][3][17][14]. PRMA protocol was proposed for packet-based data transmission over wireless networks and uses a random access protocol on the packet-transmission channel. A mobile with a periodic source successfully transmits one voice packet in an available slot to the base station reserved that slot in future frames. It prevents subsequent collisions with data packets from other terminals. At the end of its burst, the mobile releases its reservation by leaving the reserved slot empty. This leads to wastage of a slot in the transmission channel [6,7,8]. DQRUMA [2,3] supports data packet service and attempts to provide efficient bandwidth utilization by using a random access protocol only on its request-access channel. The uplink and downlink periods are configured on a slot-by-slot basis. The uplink slot comprises an extra bit called the piggyback (PGBK) bit. The base station checks this bit and updates the appropriate entry in the base station request table accordingly and then knows that mobile has more data to transmit or not. However, the disadvantage of DQRUMA protocol is that it does not make any distinction between CBR and ABR service categories. It treats both as burst traffic. Consequently, it does not consider any priority handling mechanism and source time out. ARCMA [3][16][17] is based on the DQRUMA protocol with the periodic traffic handling of PRMA. It attempts to provide better overall channel utilization by reducing contention in Request Access (RA) channel [3] with an inexplicit priority scheme. It is observed that these currently available multiple access protocols didn't deal with a full complement of multimedia signals. Further these protocols do not assign explicit priority to the different types of information, namely voice and data. Thus, there is a need for a multiple access protocol for handling full complement multimedia signals. Some of the media may be delay sensitive while others are loss sensitive. Thus the explicit priority scheme based on traffic types can improve the performance. However, the problem with priority assignment according to the type of the media type only is that some lower priority media may have to wait too long and the information may become stale. Also the stations have limited buffer space, so some of the information may be lost due to buffer overflow. Also discarding traffic coming through channels with inferior transmission quality can improve processing and transmission times for other traffics coming through channels with superior transmission qualities. Thus consideration of explicit priorities, Time to Live (TTL) of transactions and Channel State Information (CSI) parameters may improve the situation.

In this paper such a protocol for multimedia transmission called Advanced Multiple Access Protocol for Multimedia Transmission Protocol (AMAPMT) is presented. Performance of this protocol is evaluated in terms packet loss ratio (PLR), mean packet transfer delay (MPTD) and throughput under various traffic loads with constant and exponential distributions with various mean arrival rates and transaction sizes. The protocol uses parameters like time to live (TTL) of transactions, priority of individual medium and relevant channel state information (CSI) in this order to grant permission to the sources to transmit.

Performance is evaluated by implementing a simulation test bed using OPNET simulation software package. In this test bed a wireless network consisting of a selectable number of source stations, mobiles and a base station has been implemented [11]. In this test bed one can implement wireless multimedia communications with selectable distributions and mean values of transaction size, distributions and mean values of inter-arrival time of transactions, time-to-live, buffer size and a weighted round robin queuing model with selectable weights [11]. The performance of the AMAPMT protocol is evaluated by using this test bed. Results for PLR,

MPTD and throughput are obtained and presented in tabular and graphical forms for various combinations of the above-mentioned parameters [11].

Performance of this protocol is compared to that of a currently available multiple access protocol called Adaptive Request Channel Multiple Access (ARCMA). AMAPMT protocol is shown to out perform ARCMA protocol.

## 2. SERVICE TYPES CONSTITUTING MULTIMEDIA SIGNAL

A multimedia signal may consist of some or all of the five service categories: constant bit rate (CBR), real-time variable bit rate (RT-VBR), non-real-time variable bit rate (NRT-VBR), available bit rate (ABR) and unspecified bit rate (UBR).  Constant bit rate (CBR) services generate output at a constant bit rate, requires time synchronization between the traffic source and destination, require predictable response time and a static amount of bandwidth for the lifetime of a connection with low latency. Examples of such services are digital Voice and Video applications. Real-Time Variable Bit Rate (RT-VBR) services, e.g., compressed video stream or mobile Internet access application, generate information at a rate that is variable with time, requires time synchronization between the traffic source and destination require the delay to be less than a specified maximum value and a variable amount of bandwidth.  Non-Real-Time Variable Bit Rate (NRT-VBR) services that include File transfer, image and fixed Internet access applications, generate variable bit rate traffic for which there is no inherent requirement on time synchronization between the traffic source and destination and require no guaranteed delay bound.  Available Bit Rate (ABR) is a best effort service [12]. Neither data rate nor delay is guaranteed. The minimum and maximum rates are guaranteed, as is a bound on packet loss rate.  This service includes data and allows wireless system to fill its channels to the maximum capacity when CBR or VBR traffic is low.  Unspecified Bit Rate (UBR) service category is similar to NRT-VBR without guaranteed minimum rate or bound on the packet loss rate.  It is used for connections that transport variable bit rate traffic for which there is no requirement on time synchronization between the traffic source and destination e.g., file transfer, back up transfer and email without delay guarantee.

## 3. WIRELESS NETWORK

### 3.1  Introduction

A typical Wireless network consists of a base station, a number of mobiles and source stations. Each mobile is connected to a number of source stations.  The source stations may be of different types, e.g., voice stations, data stations, Ftp stations and Email stations.  In this network the source stations generate and save information (voice, video, data, Ftp, email).  The source stations send Request Access (RA) packets to the relevant mobiles to ask permission to send information.  The mobiles forward these requests to the base station.  The base station considers all such requests for a frame time and grants permission to transmit information, according to some multiple access protocol, to the source stations via relevant mobiles. Performance of the protocol is measured in terms of some desired qualities of service.

### 3.2  Available Multiple Access Protocols, Their Applicability and Shortcomings

Three multiple access protocols are available.  These are Packet Reservation Multiple Access (PRMA), Distributed Queuing Request Update Multiple Access (DQRUMA) and Adaptive Request Channel Multiple Access (ARCMA) [1, 2, 3, and 6].

The currently available multiple access protocols deal with only voice and data and not a full complement of multimedia signals. Further these protocols do not assign explicit priority to the different types of information, namely voice and data. Thus, there is a need for a multiple access protocol for handling full complement (CBR, RT-VBR, NRT-VBR, ABR and UBR) of multimedia signals [13]. Some of the media may be delay sensitive while others are loss sensitive. Thus the multiple access protocol should treat these media in a fair prioritized way. An explicit priority scheme based on traffic types can improve the performance. However, the problem with priority assignment according to the type of the media type only is that some lower priority media may have to wait too long and the information may become stale. Also the stations have limited buffer space, so some of the information may be lost due to buffer overflow. Also discarding traffic coming through channels with inferior transmission quality can improve processing and transmission times for other traffics coming through channels with superior transmission qualities. Thus consideration of explicit priorities, Time to Live (TTL) of transactions and Channel State Information (CSI) parameters may improve the situation. Such a protocol that assigns priorities on the basis of TTL, explicit priorities for the various media and CSI is proposed and evaluated in the section 4.

# 4. THE ADVANCED MULTIPLE ACCESS PROTOCOL FOR MULTIMEDIA TRANSMISSION (AMAPMT)

## 4.1   Introduction

In this paper a new multiple access protocol for wireless network called Advanced Multiple Access Protocol for Multimedia Transmission (AMAPMT) is developed and its performance is evaluated. This protocol handles the full complement of multimedia signals and uses the time to live (TTL) parameter and explicit priority assignments to media types to improve its performance and to treat the various types of media fairly. It also uses channel state information (CSI) (e.g., bit error rate) in assigning access to media. Media received over a channel with unacceptably low channel state information (e.g., high BER) are denied access.

## 4.2   The Principle of Operation of AMAPMT Protocol

In this protocol, the stations generate and save information (voice, video, data, Ftp, email). The stations send Request Access (RA) packets to the relevant mobiles to ask permission to send information. The mobiles forward these RA packets to the base station over a reservation channel using some multiple access protocol as ALOHA. The base station acknowledges the requests, saves the RA packets for a frame time. The Request Access (RA) packets contain information on the source address, destination address, media type, bit rate, time-to-live, CSI and the requested quality-of-service (QoS) of the traffic to be transmitted. The base station processes the RA packets and grants permission to the relevant stations to transmit in a fair queuing basis using the parameters contained in the RA packets in the order of TTL, CSI and traffic type. The operation of AMAPMT protocol is explained in the flow charts in section 4.3.

## 4.3   The Flow Charts of Operation of AMAPMT Protocol

The flow charts in figures 4.1 through 4.5 describe the various aspects of operation of the AMAPMT multiple access protocol.

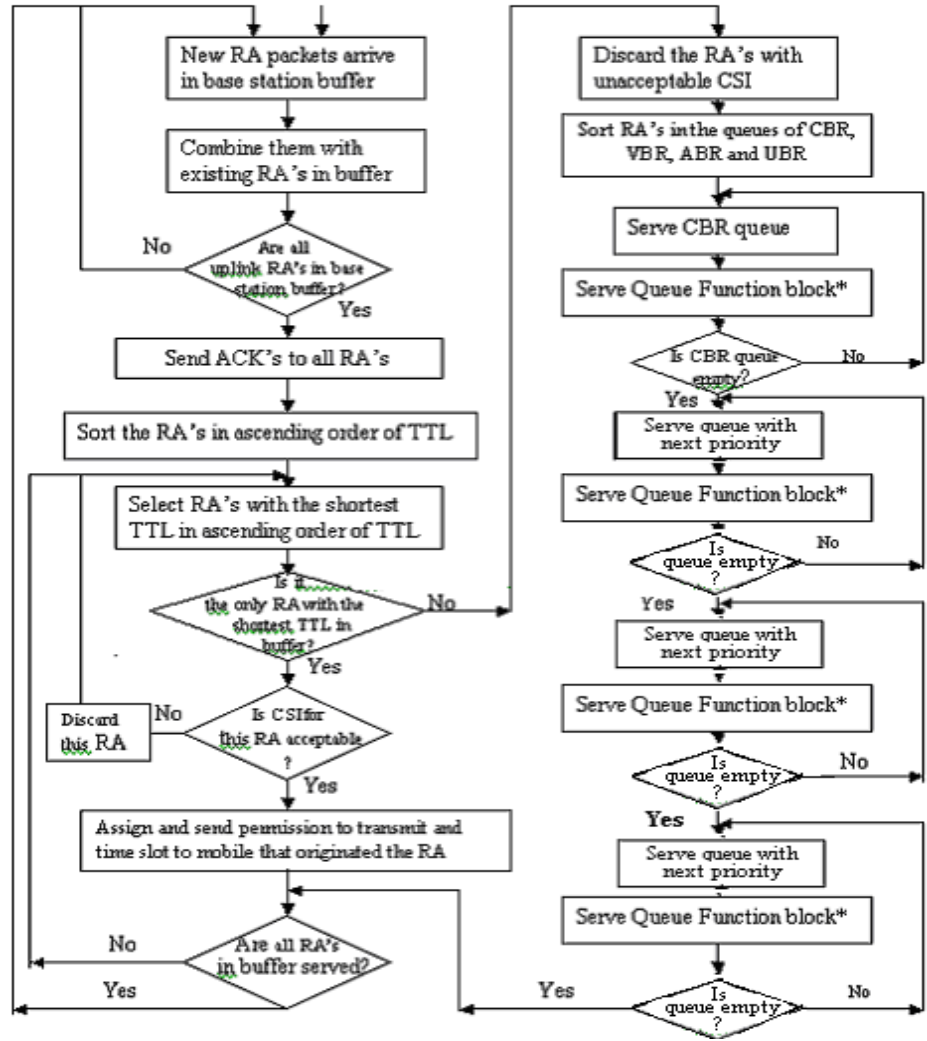

Figure 4.1. Protocol for handling of RA packets at the base station

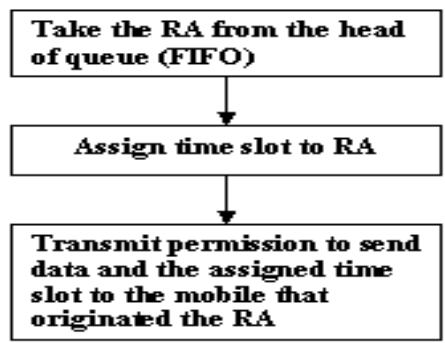

Figure 4.2. The Serve Queue Function block at the base station

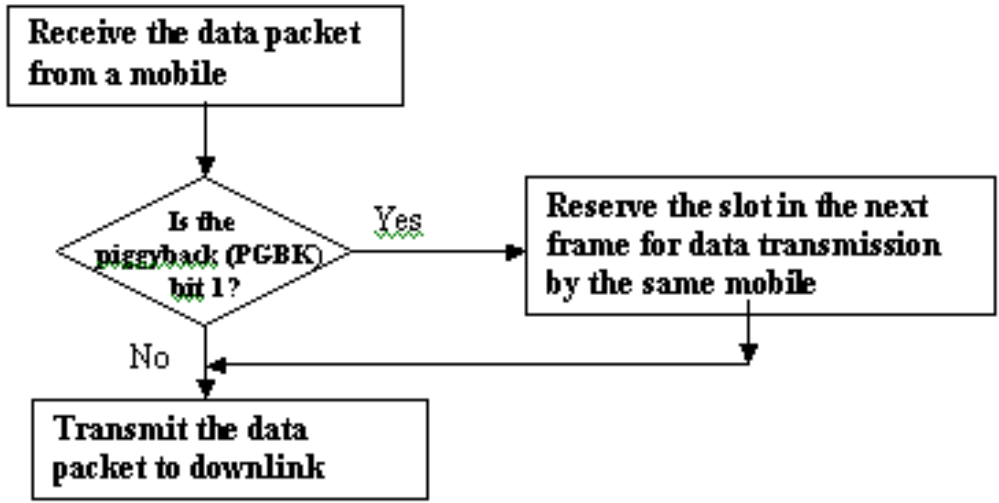

Figure 4.3. Protocol for the handling of data packets at the base station

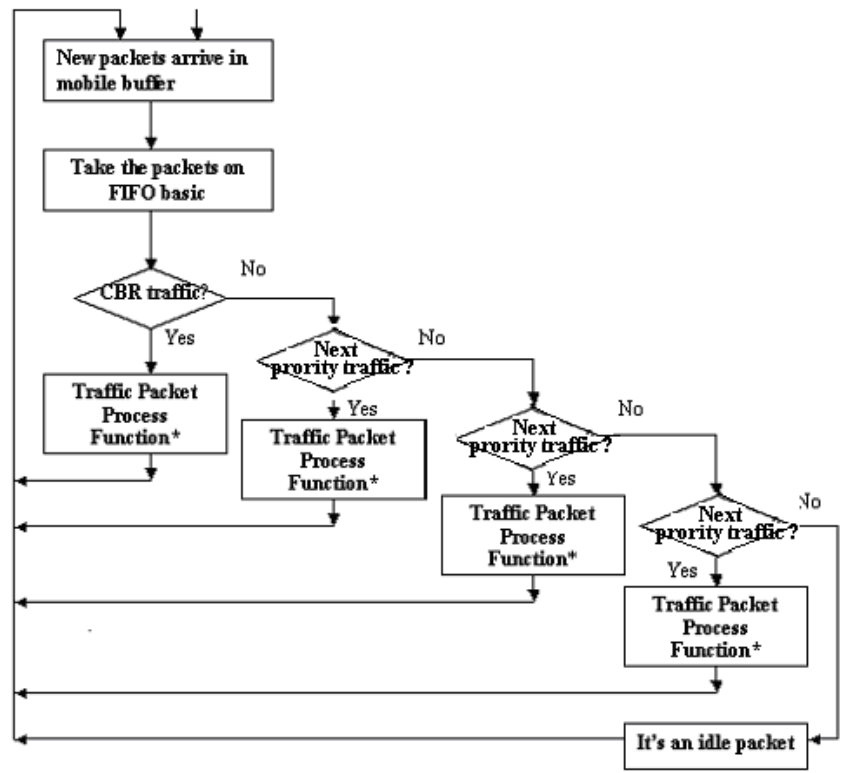

* Traffic Packet Process Function is shown Figure 4.5

Figure 4.4. Protocol at the Mobiles for handling RA packets from stations

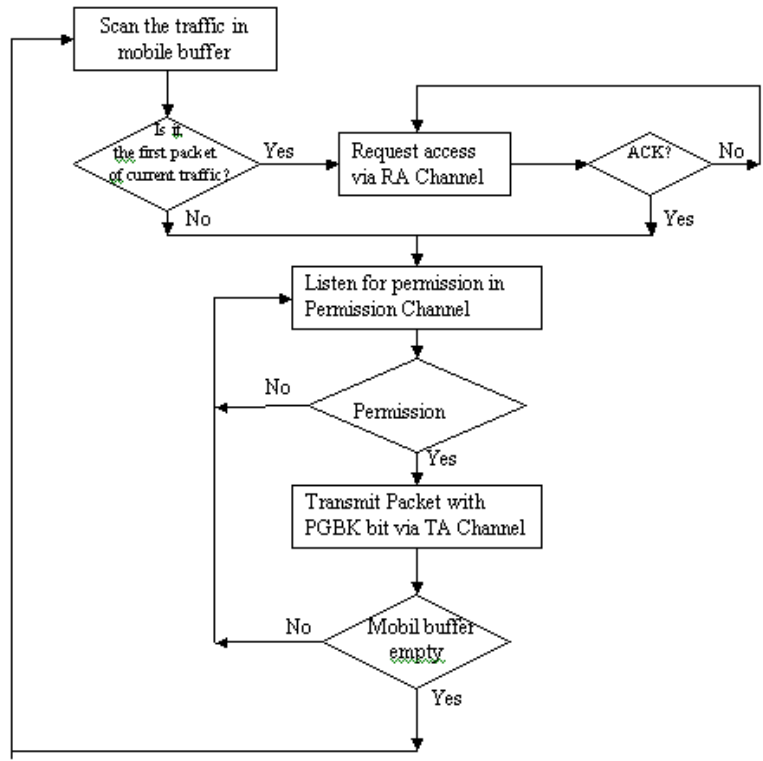

PGBK ------ Piggyback
TA -------Transmission Access
ACK ------Acknowledgement

Figure 4.5. Data Packet Process Function at the Mobiles

# 5. SIMULATION AND PERFORMANCE EVALUATION OF AMAPMT PROTOCOL

Performance of the AMAPMT protocol is evaluated by using a simulation model of a typical wireless network developed by using OPNET simulation software package [11].

## 5.1 Description of The Simulation Model

### 5.1.1 Introduction of the simulation model

The wireless network that is simulated consists of a base station, five mobiles, and four stations per mobile. The five traffic types namely Voice, Video, Data, Ftp and Email are simulated in OPNET by using CBR for the Voice, RT-VBR for Video, NRT-VBR for Ftp, ABR for data traffic and UBR for Email. The performance of the protocol is evaluated under various combinations of operational conditions of relative priorities of various media, time-to-live (TTL) and channel state information (CSI) parameters. In each case, three performance metrics, namely, Packet Loss Ratio (PLR) and Mean Packet Transmission Delay (MPTD) are obtained and compared.

### 5.1.2 Architecture and operation of the simulation model

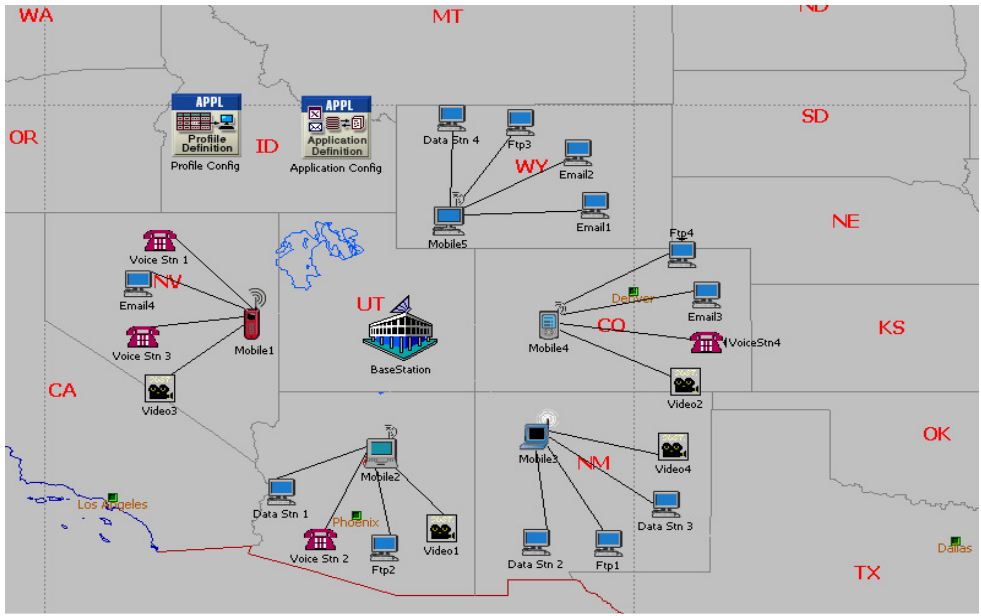

Figure 5.1 describes the overall simulation model of the network

The network model consists of a base station, five mobiles and four source stations per mobile. The source stations are of different types, e.g., voice stations, data stations, Ftp stations and Email stations. In this network the source stations generate and save information (voice, video, data, Ftp, email). The stations send Request Access (RA) packets to the relevant mobiles to ask permission to send information. The mobiles forward these requests to the base station. The base station considers all such requests for a frame time and grants permission to the stations to transmit information according to the AMAPMT multiple access protocol. A number of simulations are run and results are collected to evaluate the performance of the protocol in terms of Packet Loss Ratio (PLR) and Mean Packet Transmission Delay (MPTD) [14][15]. The following figure describes the functional model showing various components of the simulated model of the network

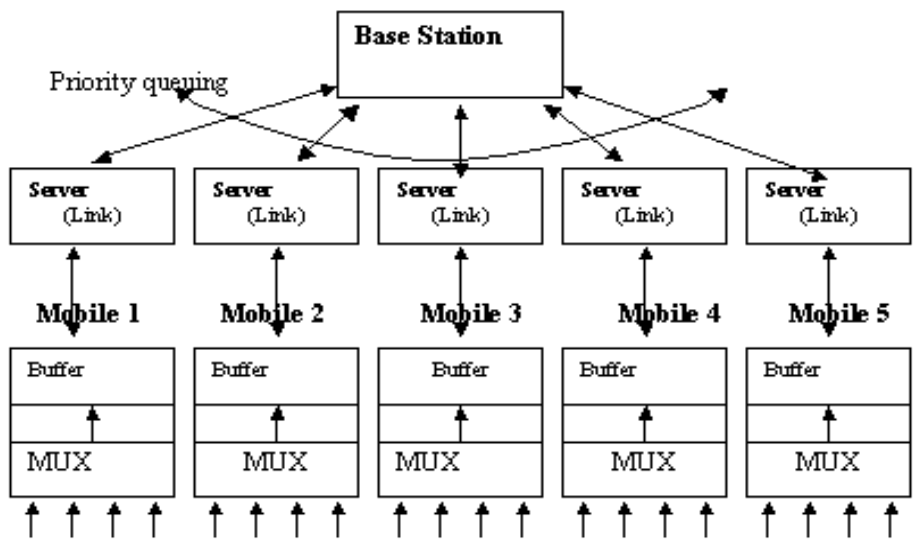

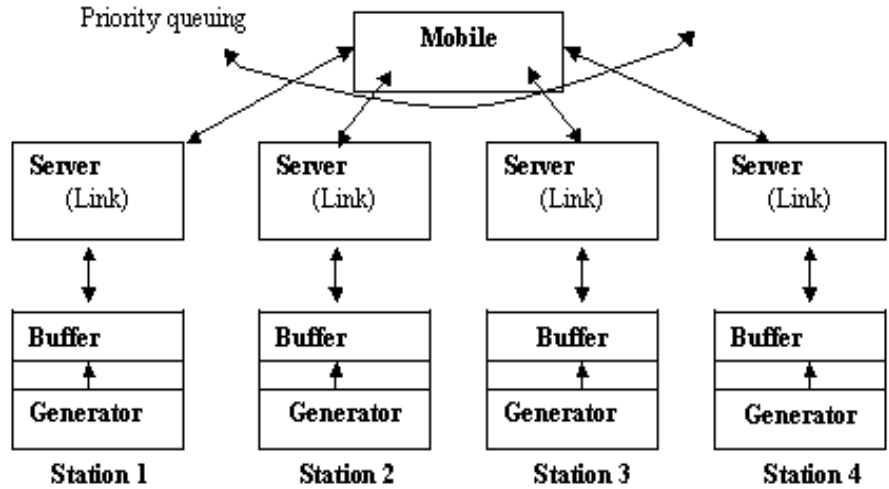

Figure 5.2   The functional block diagram of the simulated Wireless network

## 5.2    Simulation Results for Performance Evaluation of AMAPMT Protocol

### 5.2.1 Introduction

Simulation results in terms of packet loss ratio (PLR), mean packet transfer delay (MPTD) and throughput are obtained and presented for the following combinations of different relative priorities of the various media, different TTL values of transactions and channel state information (CSI) for constant and exponential distributions for source generation rates(per station).

a.  The sources per station have same priority, same TTL and same CSI (BER=1E-06 or 1E-12)
b.  The sources per station have different priorities, same TTL and same CSI (BER=1E-06 or 1E-12) )
c.  The sources per station have same priority, different TTL's and same CSI (BER=1E-06)
d.  The sources per station have different priorities, different TTL's and same CSI (BER=1E-06 or 1E-12))

The sources generate packets at a rate whose distribution constant or exponential with mean rates of up to 250,000 packets per second. The buffer size at the source stations is set up as 512 Kbytes. The buffer size at the mobile stations and the base station are assumed to be unlimited [11].  The link transmission rate is 1.54 Mbps. The channel state information (CSI) is taken as bit error ratios (BER) of 1E-06 or 1E-12. The utilization factor $\rho$= .259, obtained as the ratio of the average generation rate of 50 Kbytes per second and service rate of 1.544 Mbps, is used in all cases. The various actual values of TTL, priorities and CSI used in the simulation are shown in the following tables 6.1 through 6.4 corresponding to the combinations shown above [11].

Table 5.1   Parameter values for same TTL, same Priorities and same CSI case for sources per station

| Station # | | 1 | 2 | 3 | 4 | All Stations | All Stations |
|---|---|---|---|---|---|---|---|
| | | TTL | TTL | TTL | TTL | PRIORITY | CSI (BER) |
| CBR | VOICE | 10 | 10 | 10 | 10 | Low Latency | $10^{-6}$ |
| RT-VBR | VIDEO | 10 | 10 | 10 | 10 | 16 | $10^{-6}$ |
| NRT-VBR | FTP | 10 | 10 | 10 | 10 | 16 | $10^{-6}$ |
| ABR | DATA | 10 | 10 | 10 | 10 | 16 | $10^{-6}$ |
| UBR | EMAIL | 10 | 10 | 10 | 10 | 16 | $10^{-6}$ |

Table 5.2   Parameter values for different Priorities, same TTL and same CSI case for sources per station

| Station # | | 1 | 2 | 3 | 4 | All Stations | All Stations |
|---|---|---|---|---|---|---|---|
| | | TTL | TTL | TTL | TTL | PRIORITY | CSI (BER) |
| CBR | VOICE | 10 | 10 | 10 | 10 | Low Latency | $10^{-6}$ |
| RT-VBR | VIDEO | 10 | 10 | 10 | 10 | 16 | $10^{-6}$ |
| NRT-VBR | FTP | 10 | 10 | 10 | 10 | 8 | $10^{-6}$ |
| ABR | DATA | 10 | 10 | 10 | 10 | 4 | $10^{-6}$ |
| | UBR | 10 | 10 | 10 | 10 | 2 | $10^{-6}$ |
| | | | | | | | |

Table 5.3   Parameter values for different TTL, same Priorities and same CSI case

| Station # | | 1 | 2 | 3 | 4 | All Stations | All Stations |
|---|---|---|---|---|---|---|---|
| | | TTL | TTL | TTL | TTL | PRIORITY | CSI (BER) |
| CBR | VOICE | 10 | 35 | 60 | 85 | Low Latency | $10^{-6}$ |
| RT-VBR | VIDEO | 15 | 40 | 65 | 90 | 16 | $10^{-6}$ |
| NRT-VBR | FTP | 20 | 45 | 70 | 95 | 16 | $10^{-6}$ |
| ABR | DATA | 25 | 50 | 75 | 100 | 16 | $10^{-6}$ |
| UBR | EMAIL | 30 | 55 | 80 | 105 | 16 | $10^{-6}$ |

Table5.4 Parameter values for different TTL, different Priorities and same CSI case

| Station # | | 1 | 2 | 3 | 4 | All Stations | All Stations |
|---|---|---|---|---|---|---|---|
| | | TTL | TTL | TTL | TTL | PRIORITY | CSI (BER) |
| CBR | VOICE | 10 | 35 | 60 | 85 | Low Latency | $10^{-6}$ |
| RT-VBR | VIDEO | 15 | 40 | 65 | 90 | 16 | $10^{-6}$ |
| NRT-VBR | FTP | 20 | 45 | 70 | 95 | 8 | $10^{-6}$ |
| ABR | DATA | 25 | 50 | 75 | 100 | 4 | $10^{-6}$ |
| UBR | EMAIL | 30 | 55 | 80 | 105 | 2 | $10^{-6}$ |

## 5.2.2 Simulation results

## 5.2.2.1Results for sources with constant source generation rate distribution and BER= 1E-06

The results in terms of PLR, MPTD and Throughput for Utilization factor ρ= .259, constant source generation rate distribution and BER=1E-06 is shown in the form of graphs in the following Figures 5.3 through 5.5.

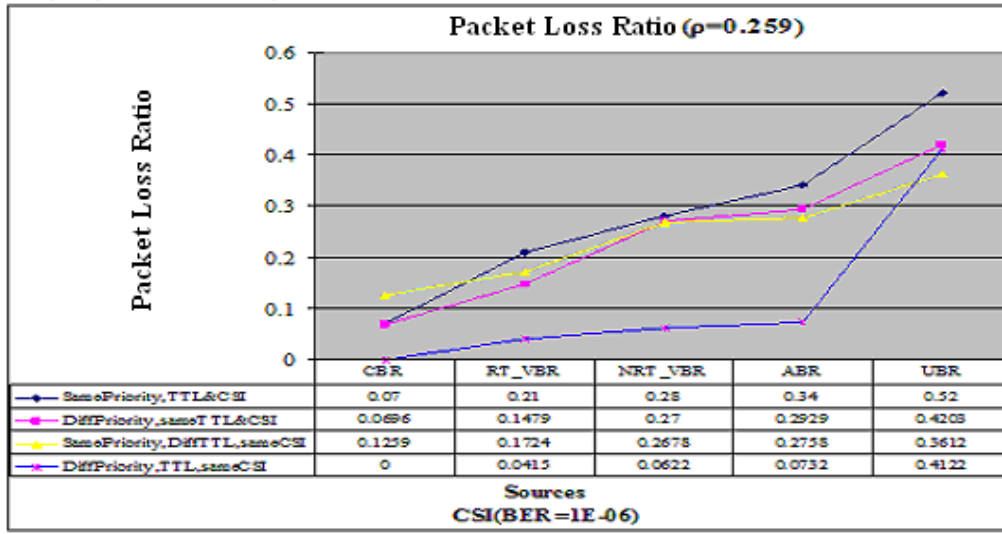

Figure 5.3 Comparison of PLR for a number of combinations of priority, TTL and CSI

It is observed from Figure 5.3 that the PLR is the highest for all media except CBR when priorities and TTL's are not considered in granting permission to transmit as in the case of ARCMA protocol. However, the PLR is slightly reduced in the cases when only priorities but not TTL's and when only TTL's and not priorities are considered as proposed in the AMAPMT protocol. Finally, the PLR values are substantially reduced for all the media when priorities and TTL's are simultaneously considered as proposed in the AMAPMT protocol. Thus the AMAPMT protocol can be used to substantially improve the PLR performance for the various components of multimedia signals.

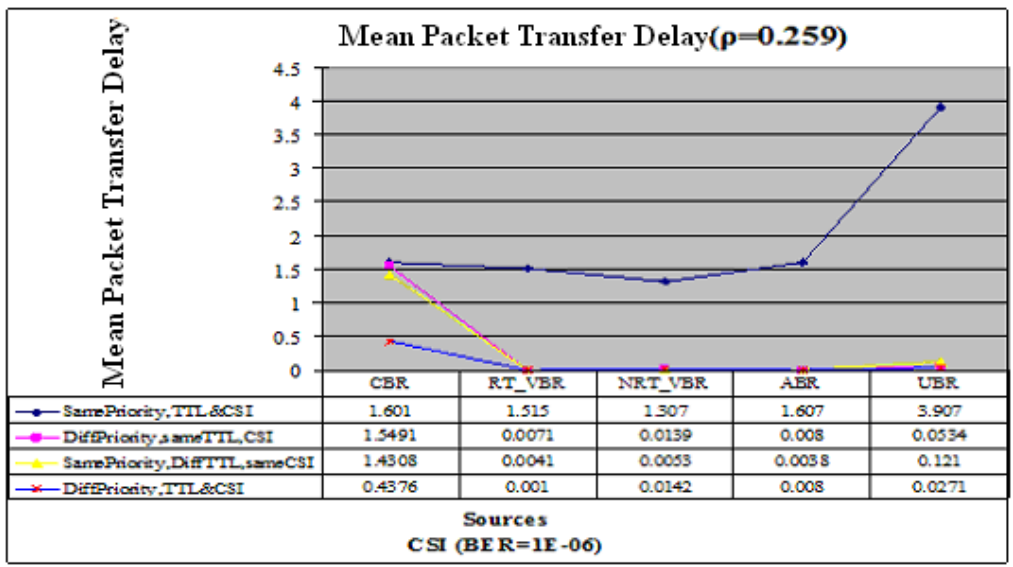

Figure 5.4 Comparison of MPTD for a number of combinations of priority, TTL and CSI

It is observed from Figure 5.4 that the MPTD is the highest for all media when priorities and TTL's are not considered in assigning permission to transmit as in the case of ARCMA protocol. However, the MPTD is considerably reduced in the cases when only priorities but not TTL's and when only TTL's and not priorities and both priorities and TTL's are considered as proposed in the AMAPMT protocol. Thus the AMAPMT protocol can be used to substantially improve the MPTD performance for the various components of multimedia signals.

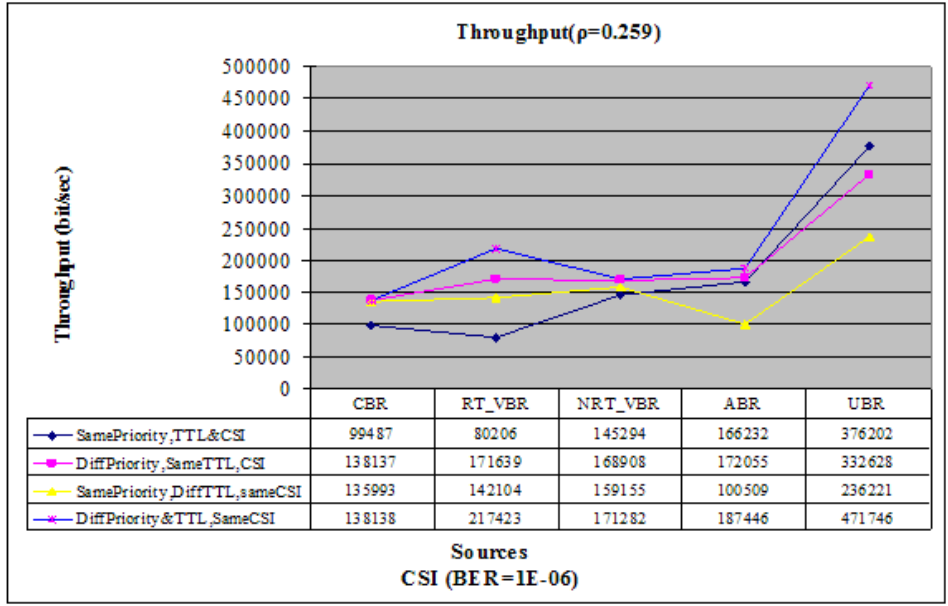

Figure 5.5   Comparison of Throughput for a number of combinations of priority, TTL and CSI as indicated on the above graph

It is observed from Figure 5.5 that the throughput is the highest for all media when priorities and TTL's are simultaneously considered in assigning permission to transmit, as proposed in the AMAPMT protocol. Thus the AMAPMT protocol can be used to improve the throughput performance for the various components of multimedia signals.

### 5.2.2.2  Results for constant source generation rate distribution and BER= 1E-12

The results in terms of PLR, MPTD and Throughput for utilization factor ρ= .259, constant source generation rate distribution and BER = 1E-12 are shown in the form of graphs in the following Figures 5.6 through 5.8.

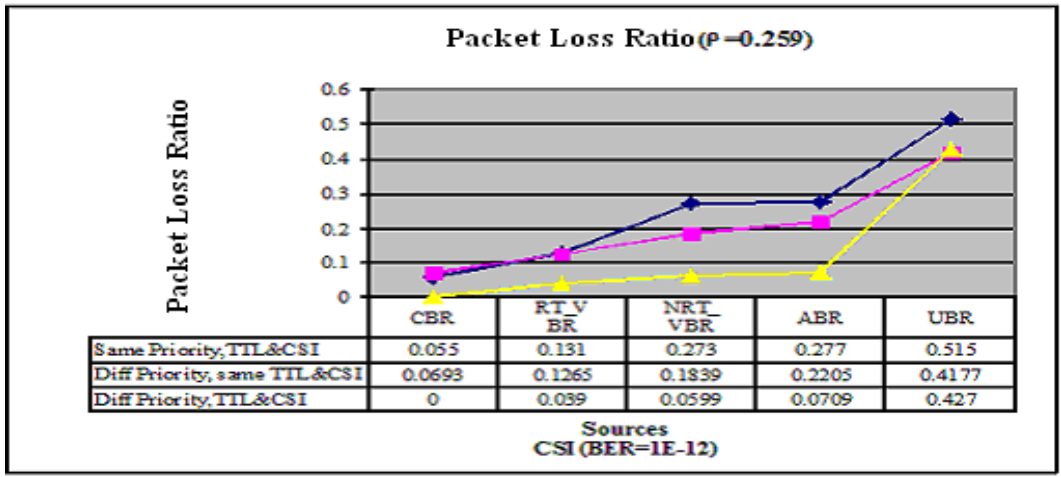

| | CBR | RT_V BR | NRT_VBR | ABR | UBR |
|---|---|---|---|---|---|
| Same Priority,TTL&CSI | 0.055 | 0.131 | 0.273 | 0.277 | 0.515 |
| Diff Priority, same TTL&CSI | 0.0693 | 0.1265 | 0.1839 | 0.2205 | 0.4177 |
| Diff Priority,TTL&CSI | 0 | 0.039 | 0.0599 | 0.0709 | 0.427 |

Figure 5.6        Comparison of Throughput for a number of combinations of priority, TTL and CSI as indicated on the above graph

It is observed from Figure 5.6 that the PLR is the highest for all media except CBR when priorities and TTL's are not considered in granting permission to transmit as in the case of ARCMA protocol. However, the PLR is slightly reduced in the case when only priorities but not TTL's are considered as proposed in the AMAPMT protocol. Finally, the PLR values are substantially reduced for all the media when priorities and TTL's are simultaneously considered as proposed in the AMAPMT protocol. Thus the AMAPMT protocol can be used to substantially improve the PLR performance for the various components of multimedia signals.

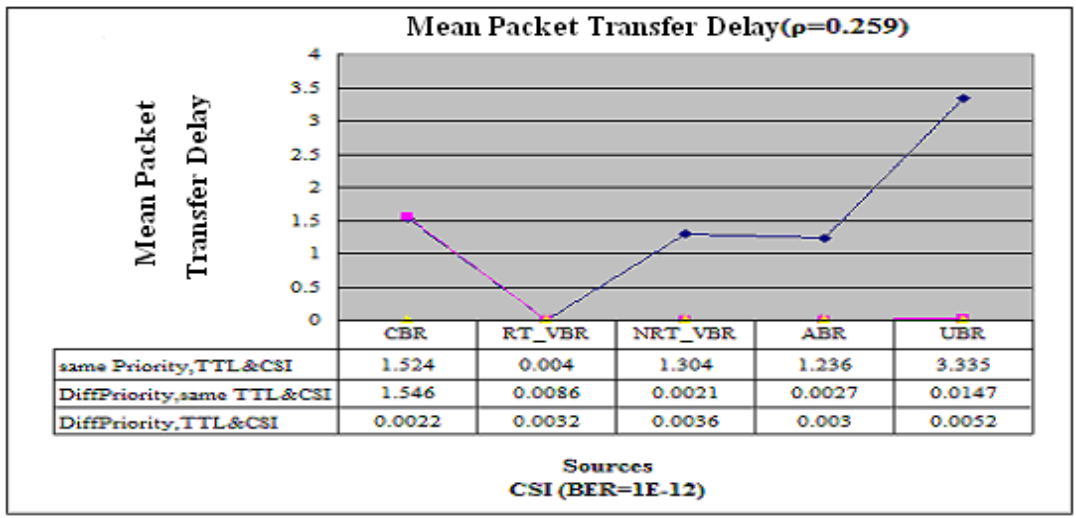

| | CBR | RT_VBR | NRT_VBR | ABR | UBR |
|---|---|---|---|---|---|
| same Priority,TTL&CSI | 1.524 | 0.004 | 1.304 | 1.236 | 3.335 |
| DiffPriority,same TTL&CSI | 1.546 | 0.0086 | 0.0021 | 0.0027 | 0.0147 |
| DiffPriority,TTL&CSI | 0.0022 | 0.0032 | 0.0036 | 0.003 | 0.0052 |

Figure 5.7 Comparison of MPTD for a number of combinations of priority, TTL and CSI

It is observed from Figure 5.3 that the MPTD is the highest for all media when priorities and TTL's are not considered in assigning permission to transmit, as in the case of ARCMA protocol. However, the MPTD is reduced in the case when only priorities but not TTL's are considered as proposed in the AMAPMT protocol. Finally, the PLR values are substantially reduced for all the media when priorities and TTL's are simultaneously considered as proposed in the AMAPMT protocol. Thus the AMAPMT protocol can be used to substantially improve the PLR performance for the various components of multimedia signals.

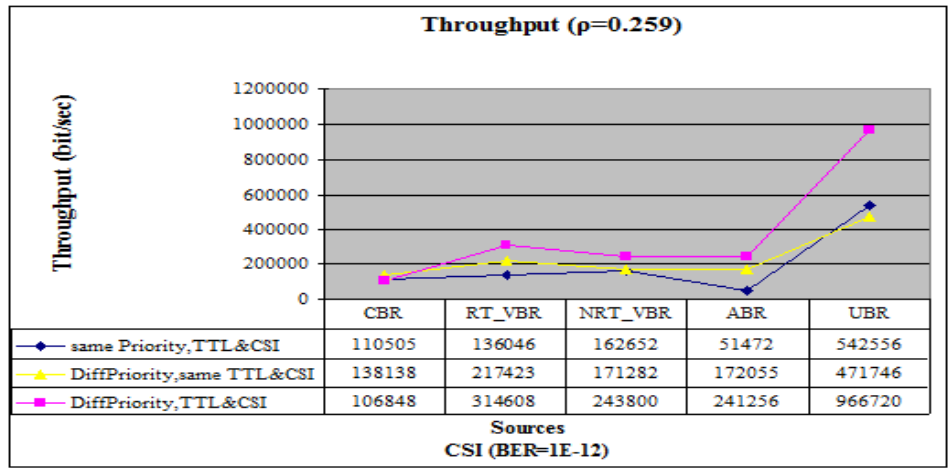

Figure 5.8  Comparison of Throughput for a number of combinations of priority, TTL and CSI as indicated on the above graph

It is observed from Figure 5.8 that the throughput is the highest for all media when priorities and TTL's are simultaneously considered in assigning permission to transmit, as proposed in the AMAPMT protocol. Thus the AMAPMT protocol can be used to improve the throughput performance for the various components of multimedia signals.

### 5.2.2.3  Comparison of results for sources with exponential source generation rate distribution and BER = 1E-06

The results in terms of PLR, MPTD and Throughput for utilization factor $\rho$= .259, exponential source generation rate distribution and BER = 1E-06 are shown in the form of graphs in the following Figures 5.9 through 5.11.

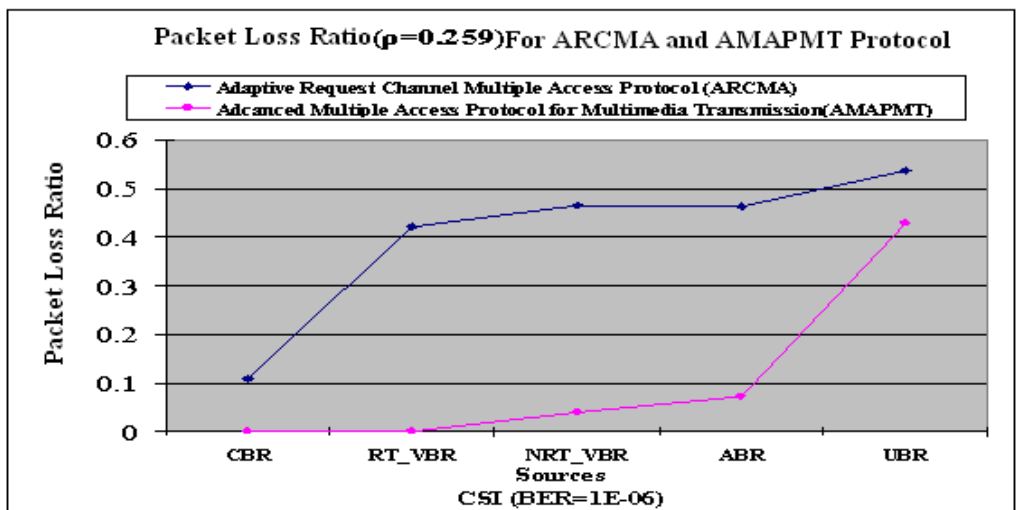

Figure 5.9   Comparison of PLR for a number of combinations of priority, TTL and CSI as indicated on the above graph

It is observed from Figure 5.9 that the PLR is the highest for most media when priorities and TTL's are not considered in granting permission to transmit as in the case of ARCMA protocol. However, the PLR values are substantially reduced for most media when priorities and TTL's are simultaneously considered as proposed in the AMAPMT protocol. Thus the AMAPMT protocol can be used to substantially improve the PLR performance for most components of multimedia signals.

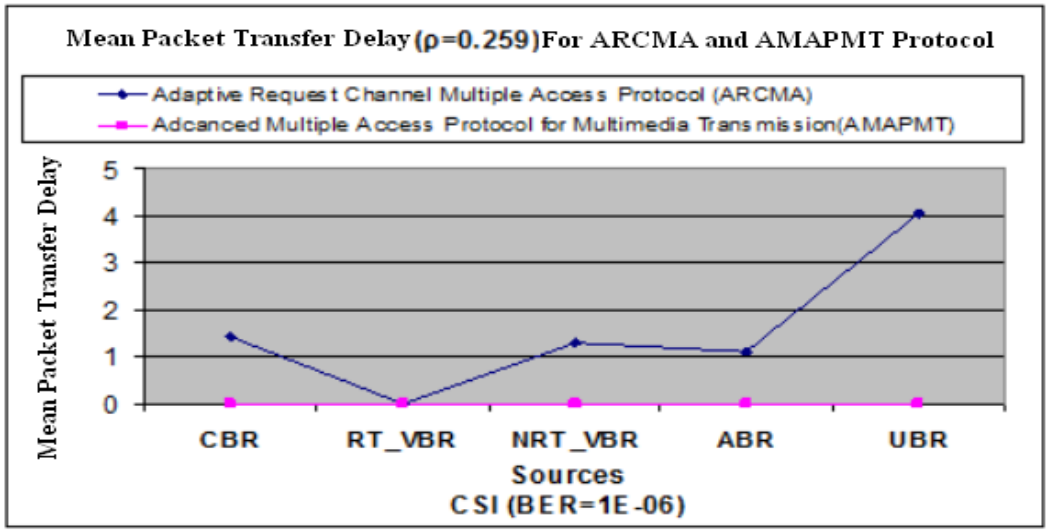

Figure 5.10    Comparison of MPTD for a number of combinations of priority, TTL and CSI as indicated on the above graph

It is observed from Figure 5.10 that the MPTD is the highest for all media when priorities and TTL's are not considered in assigning permission to transmit, as in the case of ARCMA protocol. However, the MPTD values are substantially reduced for all the media when priorities and TTL's are simultaneously considered as proposed in the AMAPMT protocol. Thus the AMAPMT protocol can be used to substantially improve the MPTD performance for the various components of multimedia signals.

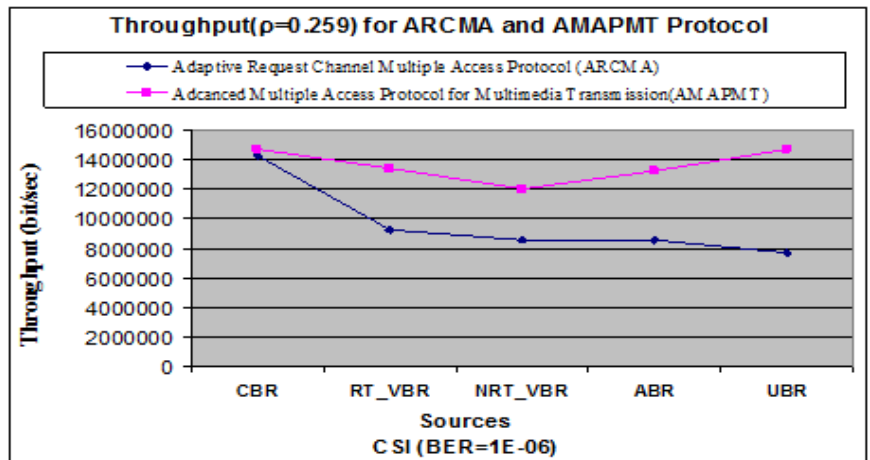

Figure 5.11 Comparison of Throughput for a number of combinations of priority, TTL and CSI as indicated on the above graph

It is observed from Figure 5.11 that the throughput is the highest for most media when priorities and TTL's are simultaneously considered in assigning permission to transmit, as proposed in the AMAPMT protocol. Thus the AMAPMT protocol can be used to improve the throughput performance for the various components of multimedia signals.

# 6. CONCLUSION

Wireless networks suitable for multimedia (Voice, Video, Ftp, Data download, Email and others) transmission are considered. A multiple access protocol called Advanced Multiple Access Protocol for Multimedia Transmission (AMAPMT) has been developed for handling full complement of multimedia traffic. A simulation model has been developed and used to evaluate PLR and MPTD and throughput for given link capacity, buffer size and input rate for selectable values of priorities of various media, TTL and CSI values. Thus the AMAPMT algorithm can be used for allocation of resources (link capacity and buffer size) at the nodes to obtain simultaneous satisfaction of prescribed end-to-end PLR, MPTD and Throughput for a given input rate.

From the results obtained it is observed that the PLR, MPTD and throughput are better for all media when priorities and TTL's are considered in assigning permission to transmit, as in the case of AMAPMT protocol than when priorities and TTL's are not considered as in ARCMA protocol. Moreover the PLR, MPTD and throughput values are substantially better for all the media when priorities and TTL's are simultaneously considered as proposed in the AMAPMT protocol. Thus the AMAPMT protocol can be used to substantially improve the PLR, MPTD and throughput performance for the various components of multimedia signals.

The results of this paper will be helpful in implementing and evaluating performance of multimedia communication over wireless networks under many selectable conditions [11]. This will help in the design of such networks and protocols. Ability to communicate with multimedia signals is becoming more and more important and is going to impact society in many ways. The advanced multiple access protocol for multimedia transmission (AMAPMT) will enhance the performance and capability of wireless networks to handle multimedia signals. This will make these networks more useful to fulfill the need of more users of multimedia signals.

## BIOGRAPHIES

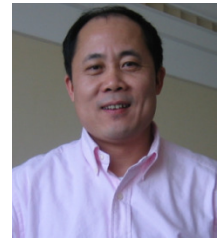

HONG YU received the B.S degree in Electrical Engineering from University of Hunan, China in 1989, the M.S. degree in Electrical Engineering from the Catholic University of America, Washington DC, in 2001, and the Ph.D. degree in Electrical Engineering from The Catholic University of America, Washington DC, in 2007. Currently, He is an Associate Professor of Electrical Engineering at Capitol College. His teaching and research areas included network communication protocol, embedded system design, optical switching application, on-line Laboratory research and renewable energy development. Dr. Yu may be reached at yhong2006@gmail.com

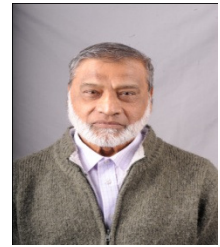

MOHAMMED AROZULLAH received his B.S degree in Physics from Rajshahi University, Bangladesh, in 1956. He earned his M. Sc degree in Physics from Dhaka University, Bangladesh, in 958 and his M.Sc degree in Electrical Engineering, University of Ottawa, Canada, 1964. He received the Ph.D degree in Electrical Engineering from University of Ottawa, Canada, 1967.
Dr. Arozullah is a Professor of Electronics Engineering and Computer Science at The Catholic University of America. His research interest include communication systems, computer communication networks, optical communications, optical networks, Internet and multimedia communications. Dr. Arozullah may be reached at arozullah@cua.edu